\documentclass[aps,pre,twocolumn,epsfig,superscriptaddress,showpacs,amsmath,amsfonts,floatfix]{revtex4}
\usepackage{graphicx}
\usepackage{dcolumn}
\usepackage{bm}

\newcommand{\CM}{\mathbb C}
\newcommand{\IM}{\mathbb I}
\newcommand{\JM}{\mathbb J}
\newcommand{\ZM}{\mathbb Z}
\newcommand{\ukr}{\hat U^{(\text{\tiny{kr}})}}
\newcommand{\udkra}{\hat U^{(\text{\tiny{dkr}})}}
\newcommand{\udkrb}{\hat U^{(\text{\tiny{ordkr}})}_{\tau}}
\newcommand{\udkrc}{\hat U^{(\text{\tiny{ordkr}})}_{2\pi p/q}}

\newcommand{\calukra}{{\cal U }^{(\text{\tiny{kr}})}(\vartheta)}
\newcommand{\calukrb}{{\cal U }^{(\text{\tiny{kr}})}_{\tilde l\tilde l'}(\vartheta)}
\newcommand{\caludkra}{{\cal U }^{(\text{\tiny{ordkr}})}(\vartheta)}
\newcommand{\caludkrb}{{\cal U }^{(\text{\tiny{ordkr}})}_{\tilde l\tilde l'}(\vartheta)}

\begin{document}
\title{Exponential Quantum Spreading in a Class of Kicked Rotor Systems near High-Order Resonances}
\author{Hailong Wang}
\affiliation{Department of Physics and Centre for Computational Science and Engineering, National University of Singapore, Singapore 117546, Republic of Singapore}
\author{Jiao Wang}
\affiliation{Department of Physics and Institute of Theoretical Physics and Astrophysics, Xiamen University, Xiamen 361005, China}
\author{Italo Guarneri}
\affiliation{Center for Nonlinear and Complex Systems, Universit\`a degli Studi dell' Insubria, Via Valleggio 11, 22100 Como, Italy}
\author{Giulio Casati}
\affiliation{Center for Nonlinear and Complex Systems, Universit\`a degli Studi dell' Insubria, Via Valleggio 11, 22100 Como, Italy}
\author{Jiangbin Gong} \email{phygj@nus.edu.sg}
\affiliation{Department of Physics and Centre for Computational Science and Engineering, National University of Singapore, Singapore 117546, Republic of Singapore}
\affiliation{NUS Graduate School for Integrative Science and Engineering, Singapore 117597, Republic of Singapore}
\date{ \today}


\begin{abstract}

Long-lasting quantum exponential spreading was recently found in a simple but very rich dynamical model, namely, an on-resonance double-kicked rotor model [J.~Wang, I.~Guarneri, G.~Casati, and J.~B.~Gong, \prl{\bf 107}, 234104 (2011)]. The underlying mechanism, unrelated to the chaotic motion in the classical limit but resting on quasi-integrable motion in a pseudoclassical limit, is identified for one special case. By presenting a detailed study of the same model, this work offers a framework to explain long-lasting quantum exponential spreading under much more general conditions. In particular,  we adopt the so-called ``spinor" representation to treat the kicked-rotor dynamics under high-order resonance conditions and then exploit the Born-Oppenheimer approximation to understand the dynamical evolution. It is found that the existence of a flat-band (or an effectively flat-band) is one important feature behind why and how the exponential dynamics emerges.  It is also found that a quantitative prediction of the exponential spreading rate based on an interesting and simple pseudoclassical map may be inaccurate.
In addition to general interests regarding the question of how exponential behavior in quantum systems may persist for a long time scale, our results should motivate further studies towards a better understanding of high-order resonance behavior in $\delta$-kicked quantum systems.

\end{abstract}


\maketitle

\section{Introduction}

In a classical chaotic system, exponential sensitivity to its initial conditions does not directly yield exponential growth in physical observables due to the complicated stretching and folding dynamics in phase space. So it sounds more unlikely to have an exponential growth in the expectation value of quantum observables for a long time scale. Indeed, after a very short time scale of dynamical evolution, even the notion of exponential sensitivity itself becomes problematic in most quantum systems. Thus, apart from prototypical  situations like  inverted harmonic oscillators \cite{barton}, long-lasting exponential quantum spreading (EQS) sounds elusive \cite{iomin}.

Recently, in a variant \cite{GongJB2011} of the kicked-rotor model \cite{Casati1995} that has been studied extensively for decades, we reported long-lasting EQS in momentum space. The found EQS is in striking contrast to other known dynamical behaviors in kicked-rotor models, such as ballistic diffusion, superdiffusion, as well as linear diffusion followed by dynamical localization \cite{Casati1995}. In particular, a double-kicked rotor model with two tunable time parameters is considered, with the first time parameter tuned precisely on the so-called main resonance. This defines the so-called on-resonance double kicked rotor model (ORDKR). The other time parameter is tuned close to the so-called anti-resonance condition \cite{Dana2005} or to high-order resonance conditions.  With the second parameter tuned close to an anti-resonance condition, the EQS mechanism is as follows. First, a quantum ORDKR, though in its deep quantum regime, can still behave remarkably close to the dynamics of a pseudo-classical limit \cite{classical}, which can be derived using the anti-resonance condition. Second, the pseudo-classical limit has one unstable fixed point, and the stable branch of the separatrix associated with the unstable fixed point can almost fully accommodate a zero-momentum initial state. Third, as time evolves, the quantum ensemble is attracted to the unstable fixed point and then exponentially repelled from the fixed point along its unstable manifold.  As also shown in Ref.~\cite{GongJB2011}, depending on the detuning from the anti-resonance condition, the time scale of this exponential behavior can be made arbitrarily long. However, Ref.~\cite{GongJB2011} was not able to explain EQS under general situations, i.e., those with the second time parameter tuned to a rather arbitrary high-order resonance condition.  It is thus still unclear what are the general and necessary conditions for EQS to occur in ORDKR.

To better understand long-lasting EQS, we carry out a detailed study and adopt a more general framework to illuminate EQS in ORDKR. Specifically, in treating the kicked-rotor dynamics under a high-order resonance condition, we adapt to ORDKR a technique of analyzing the quasi-resonant KR  dynamics \cite{Guarneri2009}, based on a spinor formulation supplemented by the Born-Oppenheimer approximation \cite{Guarneri2009}. It is found that the existence of a flat-band (or an effectively flat-band) of ORDKR at zero detuning is one important feature behind the exponential dynamics. A flat band or an almost flat band makes it possible for a very simple pseudoclassical limit to emerge.  This fact turns out to be very important to explain and understand EQS qualitatively.  However, on a quantitative level, we also observe that predictions based on a  pseudoclassical limit may not work well in accurately predicting the EQS rate. This finding is shown to have a link with the fine spectral properties of the Floquet bands of ORDKR.

The paper is organized as the follows. In Sec.~II, we first introduce necessary terms for describing a quantum kicked-rotor under resonance.  In the same section we also introduce ORDKR together with its important spectral features that can be connected with EQS. In Sec.~III, we study in detail the dynamics of ORDKR when its second time parameter is tuned close to a general high-order resonance. We discuss the dynamics induced by a small detuning from high-order resonances and then show how simple pseudoclassical limits can emerge. We then discuss in depth why EQS occurs and what should be the associated conditions. We also compare and comment on the EQS rate obtained from the pseudoclassical limits with those directly obtained from the quantum dynamics. Section IV concludes this paper.


\section{ORDKR under high-order quantum resonance conditions}


\subsection{Quantum resonances and spinor states}

In dimensionless units, the quantum map operator (Floquet operator) for a periodically kicked rotor model can be written as
\begin{equation}
\ukr = \exp\left[-i\frac{\tau}{2}{\hat l}^2\right] \exp\left[-ik\cos(\theta)\right] \;,
\label{floKR}
\end{equation}
where $\theta\in [0,2\pi)$ is an angular coordinate, $\tau$ is an effective Planck constant, $\cos(\theta)$ describes the profile of a kicking potential, $k$ is determined by the strength of the kicking potential,
and $\tau\hat l$ can be regarded as the (angular) momentum operator, whose eigenstates $|l\rangle$ with eigenvalues $\tau l$ ($l\in\ZM$) will be used as the basis states of the Hilbert space; in the $\theta$-representation, $\langle\theta|l\rangle = (2\pi)^{-1/2}\exp(il\theta)$. For later convenience, one may write $\ukr=\hat M_{\tau}\hat V$, where $\hat M_{\tau}$  and $\hat V$ respectively denote the two unitary operators on the rhs in Eq.~(\ref{floKR}).

The kicked-rotor dynamics is highly sensitive to the arithmetic  properties of $\tau$. If $\tau$ is commensurate to $2\pi$,  then the quasi-energy spectrum of $\ukr$ has a band structure , so asymptotically in time  the system's kinetic energy  $(\tau\hat l)^2/2$ grows ballistically (except in exceptional cases, which occur when all bands are flat). This is called a KR quantum resonance and  is due to translational invariance in momentum space \cite{qkrres}. Denoting $\hat T$ the shift operator in momentum space: ${\hat T}^{\dagger}\hat l\hat T=\hat l+1$, one finds that , whenever $\tau=2\pi p/q$ with $p,q$ mutually prime,
\begin{equation}
\hat T^{q\dagger} \ukr \hat T^q = (-1)^{pq}\ukr \;.
\label{tq}
\end{equation}
So $\ukr$ commutes with momentum space translations by multiples of $q$ whenever $pq$ is even, and with momentum space translations by multiples of $2q$ whenever $pq$ is odd. In both cases, the Bloch theorem is applicable in momentum space. It is sufficient to sketch this construction for the case when $pq$ is even, as adaptation to the other case is trivial.

Thanks to Eq.~(\ref{tq}), the generator of the translation ${\hat T}^q$ is conserved under the discrete time evolution generated by $\ukr$. Since this translation acts in momentum space, its generator may be dubbed quasi-position.  Using that ${\hat T}^q$ is diagonal in the $\theta$-representation, where it corresponds to multiplication by $\exp(iq\theta)$, one finds that in the $\theta$-representation quasi-position  is an eigenphase of ${\hat T}$, given by $\vartheta = q\theta\pmod{2\pi} = q\left[\theta\pmod{2\pi/q}\right]$. Conservation of quasi-position makes it convenient to study the resonant dynamics in a  representation where quasi-position, and hence ${\hat T}^q$, is diagonal. Such a representation  is provided by any eigenbasis  of ${\hat T}^q$. Straightforward calculation shows that in the momentum $l$-representation any (generalized) eigenvector $u$ of ${\hat T}^q$ associated with a quasi-position $\vartheta$ has the Bloch form:
\begin{equation}
\langle l|u\rangle = e^{-i\overline l\vartheta}f(\vartheta,\tilde l) \;,
\label{blocheigf}
\end{equation}
where $\tilde l=l\pmod{q}$, $l=\overline lq+\tilde l$,  and $f(\vartheta, \cdot)$ is an arbitrary function defined on the ``unit cell" $\{0,1,\ldots,q-1\}$ (letters marked by a tilde, like $\tilde l,\tilde l',\ldots$  will hereafter denote integers drawn from the ``unit cell"). We shall therefore introduce a representation, for which a complete set of commuting observables is provided by quasi-position $\vartheta$ along with the observable $\tilde l$  that is defined as: $\hat{\tilde l} = \sum\limits_{l=-\infty}^{\infty} \left(l\bmod{q}\right)|l\rangle\langle l|$. A basis for this representation consists of (generalized) eigenvectors  $|\vartheta,\tilde l\rangle$, ($0\leq\vartheta<2\pi)$, that are represented  in the momentum $l$-representation by :
\begin{equation}
\langle l|\vartheta,\tilde l'\rangle = (2\pi)^{-1/2}e^{-i\overline l\vartheta}\delta_{\tilde l,\tilde l'} \;.
\label{eigfs}
\end{equation}
They indeed have the Bloch form, see Eq.~(\ref{blocheigf}), with the same meaning of $\overline l$ and $\tilde l$,  and satisfy:
\begin{equation*}
\langle\vartheta,\tilde l|\vartheta',\tilde l'\rangle = \delta(\vartheta-\vartheta') \delta_{\tilde l,\tilde l'} \;,
\end{equation*}
along with the completeness relation:
\begin{equation*}
\hat{\IM} =\int_{0}^{2\pi}d\vartheta \sum_{\tilde l=0}^{q-1} |\vartheta,\tilde l\rangle\langle\vartheta,\tilde l| \;.
\end{equation*}
In this representation any rotor state $\Psi$ is represented by a wave function $\tilde\psi(\vartheta,{\tilde l}) \in L^2([0,2\pi]) \otimes \CM^q$ , that is naturally interpreted as a $q$-component spinor wave function of the coordinate $\vartheta$, and is related to the wave function $\psi(l)$ in the $l$-representation by:
\begin{equation}
\begin{split}
\tilde\psi(\vartheta,\tilde l) &= \langle\vartheta,\tilde l|\Psi\rangle = \langle\vartheta,\tilde l| \sum_{l=-\infty}^{\infty} \psi(l)|l\rangle \\
&= \langle\vartheta,\tilde l| \sum_{\overline l=-\infty}^{\infty} \sum_{\tilde l'=0}^{q-1} \psi(\overline lq+\tilde l')|\overline lq+\tilde l'\rangle \\
&= \frac{1}{\sqrt{2\pi}} \sum_{\overline l=-\infty}^{\infty} e^{i\overline l\vartheta}\psi(\overline lq+\tilde l) \;.
\end{split}
\end{equation}
It is easily seen that in the spinor $(\vartheta,\tilde l)$-representation the angular momentum $\hat l$ is represented by the operator $-iqd/d\vartheta+\tilde l$, which may be pictured as a decomposition of the angular momentum in  an ``orbital" plus a ``spin" part.

A quite similar calculation yields matrix elements of the Floquet operator in this representation:
\begin{equation}
\begin{split}
&\langle\vartheta,\tilde l|\ukr|\vartheta',\tilde l'\rangle = \delta(\vartheta-\vartheta') {\calukrb}, \\
&{\calukrb} := \sum_{\overline l=-\infty}^{\infty} \langle\tilde l|\ukr|{{\overline l}q+\tilde l'}\rangle e^{-i{\overline l}\vartheta} \;.
\end{split}
\label{matele}
\end{equation}
For any fixed quasi-position $\vartheta$, ${\calukrb}$ represents matrix elements of a $q\times q$ unitary matrix ${\calukra}$. The $q$ eigenphases of this matrix depend on $\vartheta$. As $\vartheta$ varies in $[0,2\pi)$ any such eigenphase $\Omega(\vartheta)$ is either a constant, generating a proper, infinitely  degenerate eigenvalue (a ``flat band") \cite{jiao2013,guarneriahp} in the spectrum of $\ukr$, or else it sweeps a continuous band in the spectrum of $\ukr$. In computing the matrix elements [see Eq.~(\ref{matele})], it is convenient to use  $\ukr=\hat M_{2\pi p/q}\hat V$. Matrix elements of $\hat M_{2\pi p/q}$ and $\hat V$ on the $|\vartheta,\tilde l\rangle$ basis are calculated  in Appendix~\ref{matelapp}, both for the case when $pq$ is even, and for the case when $pq$ is odd. It is found that
\begin{equation}
\begin{split}
\langle\vartheta,\tilde l|\hat M_{2\pi p/q}|\vartheta',\tilde l'\rangle &=\delta(\vartheta-\vartheta'-\pi pq) {\cal M}_{ \tilde l\tilde l'} \\
\langle\vartheta,\tilde l|\hat V|\vartheta',\tilde l'\rangle &=\delta(\vartheta-\vartheta') {\cal V}_{\tilde l\tilde l'}(\vartheta) \;.
\end{split}
\label{matmv}
\end{equation}
The matrices ${\cal M}$ and ${\cal V}(\vartheta)$ are explicitly presented in Appendix~\ref{matelapp}; the matrix ${\cal M}$ turns out to be diagonal and independent of $\vartheta$. Note that if $pq$ is an odd integer then $\hat M_{2\pi p/q}$ is not diagonal, reflecting that it is not invariant under momentum translations by $q$.

We next apply this formalism to a double-kicked rotor model \cite{Monteiro2004}, where within each period the free evolution of a rotor is interrupted twice by an external kicking potential. Again in terms of dimensionless parameters, the Floquet operator is given by
\begin{equation}
\begin{split}
\udkra =& \exp\left[-\frac{i}{2}(1-\eta)T\hat l^2\right] \exp\left[-ik\cos(\theta)\right] \\
\times& \exp\left[-\frac{i}{2}\eta T\hat l^2\right] \exp\left[-ik\cos(\theta)\right] \;,
\end{split}
\end{equation}
where $\eta T$ is the time interval between two subsequent kicks, with $0<\eta<1$.  As in previous work \cite{GongJB2008}, we further assume that the overall period $T$ of the system has been tuned to $T=4\pi$. This leads to the so-called ORDKR model, whose Floquet propagator becomes:
\begin{equation}
\begin{split}
\udkrb =& \exp\left[\frac{i}{2}\tau\hat l^2\right] \exp\left[-ik\cos(\theta)\right] \\
\times& \exp\left[-\frac{i}{2}\tau\hat l^2\right] \exp\left[-ik\cos(\theta)\right] \;,
\end{split}
\label{ordkr}
\end{equation}
where $\tau :=\eta T$ now plays the role of an effective Planck constant. Note that for $p=q=1$, $\tau=2\pi$, $\udkrb$ is equivalent to a single kicked-rotor model under the so-called anti-resonance condition \cite{Dana2005}. Using the operators $\hat M_{\tau}$ and $\hat V$ which were introduced after Eq.~(\ref {floKR}), one may write:
\begin{equation}
\udkrb = \hat M^{\dagger}_{\tau}\hat V\hat M_{\tau}\hat V \;,
\label{dkprod}
\end{equation}
whence, referring to Eq.~(\ref{tq}), it is immediately seen that whenever $\tau=2\pi p/q$ with $p,q$ coprime, $\udkrb$ commutes with translations by $q$ in momentum space, regardless of parity of $pq$; so the spinor formalism which was described above for the KR can be equally applied here. The matrix-valued function of quasi-position ${\caludkra}$ that represents $\udkrc$ in the $(\vartheta,\tilde l)$-representation is immediately found using Eq.~(\ref{matmv}) and (\ref{dkprod}) :
\begin{equation}
{\caludkra} = {\cal M}^{\dagger}{\cal V}(\vartheta-\pi pq){\cal M}{\cal V}(\vartheta) \;,
\label{dkmatprod}
\end{equation}
 where ${\cal M}, {\cal V}(\vartheta)$ are the matrices that appear in Eq.~(\ref{matmv}). The spectral analysis of the ORDKR at exact resonance, $\tau=2\pi p/q$, is based on diagonalization of the matrices ${\caludkra}$. Some results are presented in the next subsection.

\subsection{Some Spectral Properties.}
\label{SSP}

\begin{figure} 
\centering 
\includegraphics[width=\linewidth]{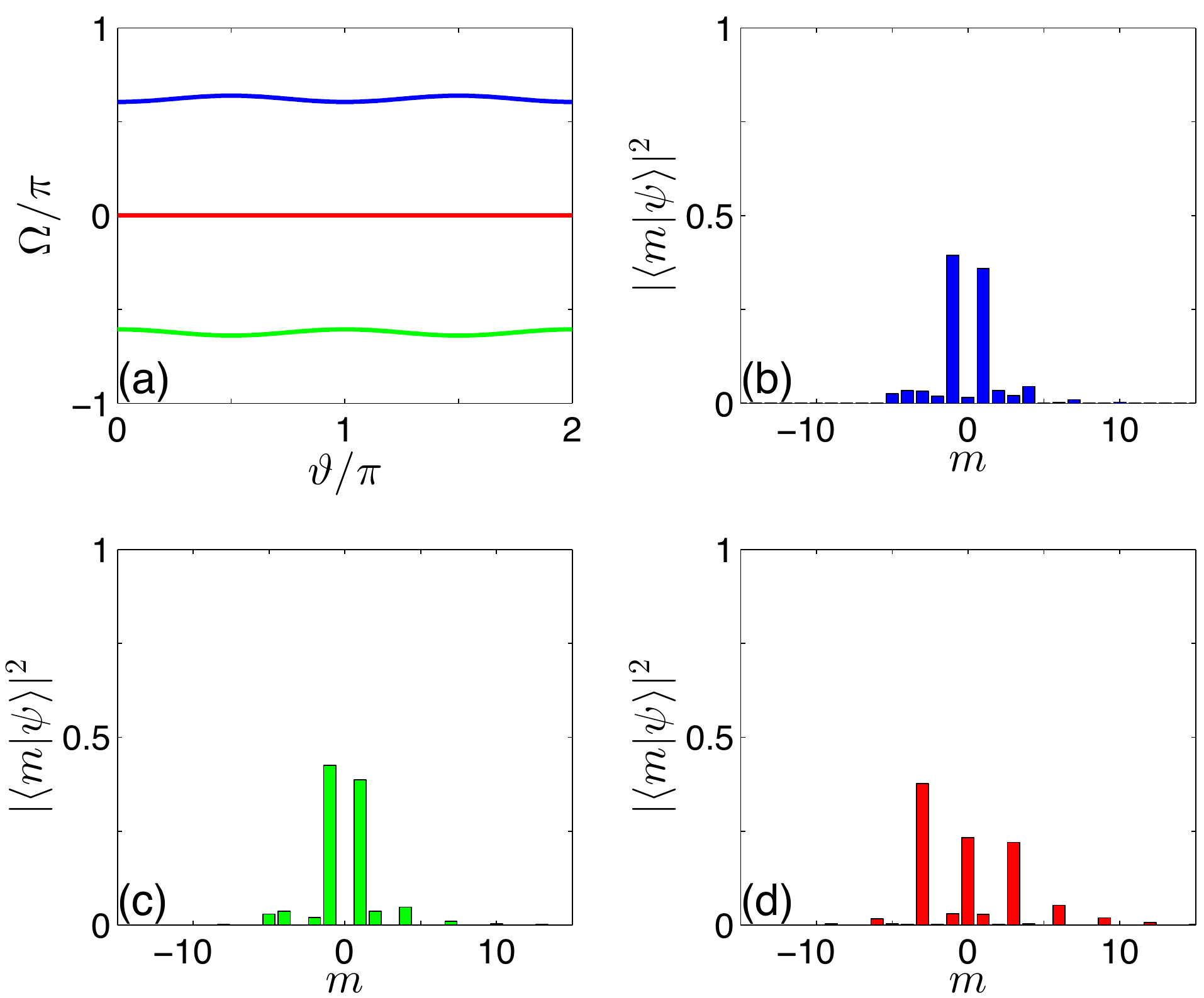}
\caption{(color online)
(a) Floquet band structure of $\udkrc$ with $p=1$, $q=3$ and $k=2.0$, shown via eigenphase $\Omega(\vartheta)$ vs $\vartheta$. Momentum distribution profiles for states prepared on bottom band (b), middle band (c), top band (d) are also shown. Here and in all other figures, all plotted quantities are in dimensionless units.}
\label{fig:rDKRM_FloSta}
\end{figure}

The band structure of $\udkrc$ has to be found numerically in general. However, in the case when $p=1$ and $q=3$, it can be explicitly computed analytically as shown in Appendix~\ref{solvablecase.apx}: a result is shown for $k=2.0$ in Fig.~\ref{fig:rDKRM_FloSta}(a). The bands are seen to be symmetric with respect to the central band, which is flat, and aligned with the zero eigenphase axis. In Appendix~\ref{flatb.apx} we show that the spectrum of $\udkrc$ exhibits exactly the same features whenever $p$ and $q$ are odd and coprime; in the following we always restrict to such cases. The dynamical properties that make the object of this paper rest on this very fact, and on a result \cite{jiao2013} that for the same values of $\tau$ the maximal bandwidth of the non-flat bands scales with the kicking strength $k$ as $k^{q+2}$, implying that under a high-order resonance condition (i.e., $q>2$), and for a sufficiently small kicking parameter $k$, effectively all the Floquet bands of $\udkrc$ will be flat. Thanks to the inversion symmetry of the spectrum
with respect to zero eigenphase \cite{jiao2013}, the eigenphases $\Omega(\vartheta)$ of the matrix ${\caludkra}$ may be labeled by a band index $\nu$ with $-q_0\leq \nu\leq q_0$, where $q=2q_0+1$, such that  $\Omega_{\nu}(\vartheta)=-\Omega_{-\nu}(\vartheta)$ ; then  symmetric bands have opposite $\nu$, and the flat band has $\nu=0$. Components of the eigenvectors of  matrix ${\caludkra}$ are likewise denoted  $u_{\tilde l}^{(\nu)}({\vartheta})$ (assuming an arbitrary $\vartheta$-dependent phase factor to be fixed by some gauge convention) . A new representation is thereby defined by quasi-position and by the quantum number $\nu$, such that basis kets
$|\vartheta,\nu\rangle$ are given in the $(\vartheta,\tilde l)$-representation by:
\begin{equation}
\langle\vartheta,\tilde l|\vartheta',\nu\rangle = \delta(\vartheta-\vartheta') u_{\tilde l}^{(\nu)}(\vartheta) \;.
\end{equation}
This may be dubbed ``the band representation".  In this representation, a state $\Psi$ is represented by a wave function:
\begin{equation}
\phi(\vartheta,\nu) := \langle\vartheta,\nu|\Psi\rangle = \sum_{\tilde l=0}^q u_{\tilde l}^{(\nu)*}(\vartheta) \tilde\psi(\vartheta,\tilde l) \;.
\end{equation}
Transformation formulae from the band representation to the momentum representation and to the coordinate representation are then easily computed in the form:
\begin{equation}
\begin{split}
\langle l|\vartheta,\nu\rangle &= (2\pi)^{-1/2} e^{-i\overline{l}\vartheta} u_{\tilde l}^{(\nu)}(\vartheta), \\
\langle\theta|\vartheta,\nu\rangle &= \delta(q\theta-\vartheta) \sum_{\tilde l=0}^{q-1} e^{i{\tilde l}\theta} u_{\tilde l}^{(\nu)}(\vartheta) \;.
\label{varthetanu}
\end{split}
\end{equation}
where as usual $l=\overline{l}q+{\tilde l}$. Under a quantum resonance condition, states prepared in a flat band subspace will stay localized, whereas states prepared in non-flat band subspaces will undergo ballistic spreading in momentum space. It will prove useful to prepare states, that are spectrally supported by the flat band alone, and in addition are well localized in momentum space around momentum $l=0$. To this end one may exploit the Mean ergodic theorem \cite{RS80} according to which, for any rotor state $|\Psi\rangle$, in the limit when $N\to+\infty$ the average
\begin{equation}
\frac1N\sum\limits_{n=0}^{N-1}\left(\udkrc\right)^n|\Psi\rangle
\label{met}
\end{equation}
tends to the projection of $|\Psi\rangle$ onto the eigenspace of $\udkrc$ which corresponds to the eigenvalue $1$, {\it i.e.} onto the flat band subspace.  Alternatively  one may compute Wannier states in momentum space, centered at $l=0$; these are equal-weight coherent superpositions of all eigenstates in a band with different choices of the Bloch phase (quasi-position) $\vartheta$. Figure \ref{fig:rDKRM_FloSta}(b),(c), and (d) show the momentum distribution profiles for such kind of states that are prepared on the bottom band, the middle band and the top band in the case $p=1,q=3$. The momentum expectation values of the states are set to  $0$, as illustrated in Fig.~\ref{fig:rDKRM_FloSta}(b)-(d).

\begin{figure} 
\centering 
\includegraphics[width=\linewidth]{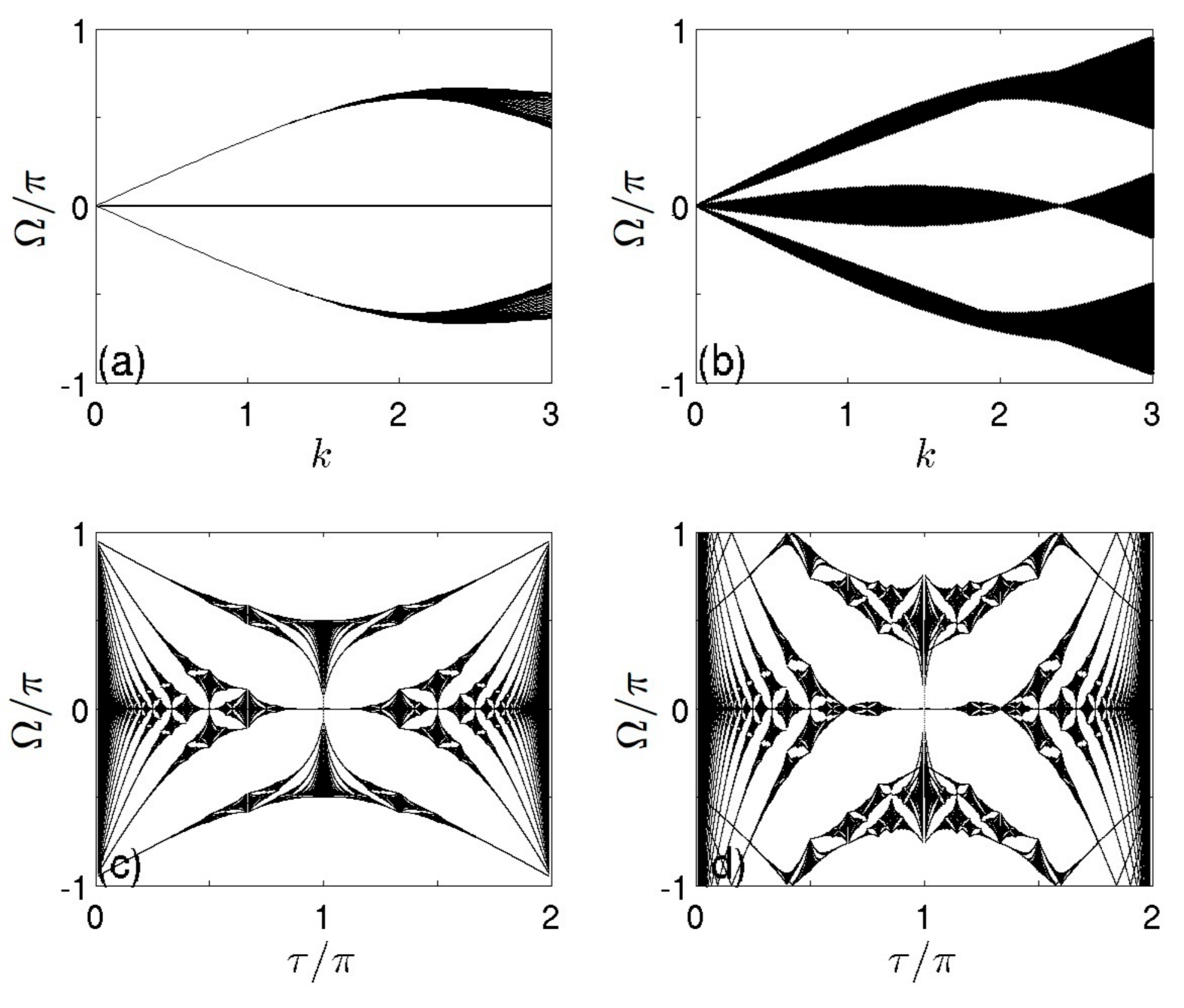}
\caption{Floquet spectrum of $\udkrb$ as a function of $k$ for $\tau=2\pi/3$ in panel (a) and for $\tau=2\pi/3+2\pi/3003$ in panel (b).  The spectrum is also collectively plotted vs a varying $\tau$ for  $k=1.5$ in panel (c) and for $k=2.4$ in panel (d).}
\label{fig:rDKRM_Flo_spectrum}
\end{figure} 
We also plot in Fig.~\ref{fig:rDKRM_Flo_spectrum}(a) the Floquet band structure vs the kicking strength parameter $k$, with $\tau=2\pi/3$. As a comparison, we show in Fig.~\ref{fig:rDKRM_Flo_spectrum}(b) the spectrum in parallel if the effective Planck constant is slightly increased to $\tau=668\pi/1001$
($=2\pi/3+2\pi/3003$, so the effective Planck constant remains a rational multiple of $2\pi$). In that case there are 1001 Floquet bands forming three clusters.  We denote $W(k)$ as the spectral range of the clusters with a kicking strength $k$. It is seen from Fig.~\ref{fig:rDKRM_Flo_spectrum}(b) that $W(k)$ is not a monotonous function of $k$. To have an overview of the spectrum , we also show for completeness the spectrum collectively as we scan $\tau$, resulting in the well-known Hofstadter's butterfly spectrum \cite{GongJB2008,GongJB2009}. The results are shown in Fig.~\ref{fig:rDKRM_Flo_spectrum}(c) for $k=1.5$ and in Fig.~\ref{fig:rDKRM_Flo_spectrum}(d) for $k=2.4$.


\section{EQS in ORDKR tuned near quantum resonance}

\subsection{Numerical results}
When setting the effective Planck constant $\tau$ of ORDKR close to a quantum resonance condition, i.e., $\tau=2\pi p/q+\epsilon$ with both $p$ and $q$ being odd integers, we found in Ref.~\cite{GongJB2011} that long-lasting EQS over many orders of magnitude of the kinetic energy occurs, with the EQS time scale increasing with a decreasing detuning parameter $\epsilon$. However, the theoretical analysis in Ref.~\cite{GongJB2011} was applicable exclusively to the case of $p=q=1$. It is still unclear what is the underlying physics behind EQS under general near-resonance cases and why we need $q$ and $p$ to be both odd integers to observe EQS.  Here we analyze the dynamics of ORDKR tuned near a general resonance condition.  Though our discussions will be as general as possible, we mainly use $\tau=2\pi/3+\epsilon$ to present explicit results.

\begin{figure} 
\centering 
\includegraphics[width=\linewidth]{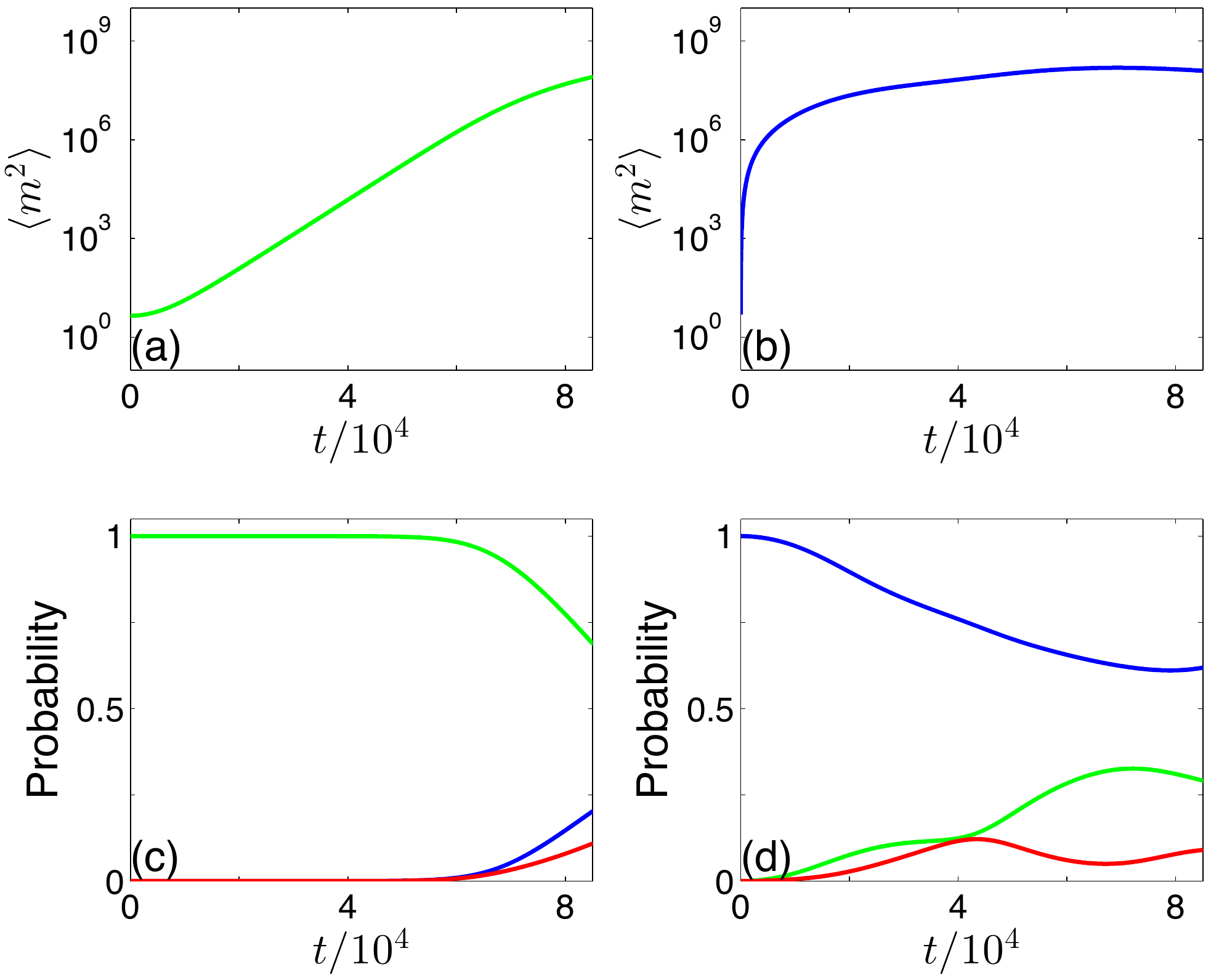}
\caption{ (color online) Panels (a) and (b) depicts the time dependence of momentum squared, with $t$ denoting the number of iterations of the ORDKR Floquet operator $\udkrb$, with $\tau=2\pi/3+\epsilon$, $k=2.0$ and $\epsilon=10^{-4}$, for initial states prepared on the middle flat band or a non-flat band of $\udkrc$ (with $p=1$ and $q=3$), respectively.  Note the log scale used for plotting the expectation value of momentum squared and that panel (a) displays a wide time window in which the plotted curve is linear, thus signaling an exponential time dependence. Panels (c) and (d) show the corresponding population dynamics, where the occupation probability is obtained by projecting the time evolving state on the three Floquet bands of $\udkrc$ (with $p=1$ and $q=3$).}
\label{fig:rDKRM_evo_1_3_k_2}
\end{figure}

Figure \ref{fig:rDKRM_evo_1_3_k_2}(a) and \ref{fig:rDKRM_evo_1_3_k_2}(b) depict the dynamics of the expectation value of momentum squared (plotted on a log scale), for an initial state prepared on the middle flat Floquet band or on the bottom non-flat band of $\udkrc$ with $p=1$, $q=3$ [throughout we all assume that such band states are localized in the neighborhood of zero momentum, see Fig.~\ref{fig:rDKRM_FloSta}(b)-(d)].  The detuning is chosen to be $\epsilon=10^{-4}$. First of all, for the flat-band initial state [panel (a)], there is an obvious (relatively wide) time window in which the plotted curve can be fitted by a straight line. This indicates EQS, which is seen to cover the expectation value of momentum squared by many orders of magnitude over a quite a long time scale.  By contrast, for the initial state prepared on a non-flat band of $\udkrc$, there is clearly no such exponential behavior.  This comparison suggests that it is important for the initial state to be placed on a flat-band of $\udkrc$ to observe EQS upon introducing a small detuning.  To further understand this, we project the wavefunction evolving in time under $\udkrb$ onto the three bands of $\udkrc$ with $p=1$ and $q=3$. We show in Fig.~\ref{fig:rDKRM_evo_1_3_k_2}(c) and (d) the time dependence of the resulting projection probabilities. In particular,  Fig.~\ref{fig:rDKRM_evo_1_3_k_2}(c) is for the flat-band initial state. There it is seen that despite the detuning $\epsilon$, the system still mainly occupies the flat-band for a very long time scale.  Interestingly, once considerable population has been transferred to the other two bands (at about $t=6\times 10^{4}$), the associated time dependence of the momentum squared shown in Fig.~\ref{fig:rDKRM_evo_1_3_k_2}(a) also starts to deviate significantly from an exponential law.  Figure \ref{fig:rDKRM_evo_1_3_k_2}(d) shows the other case, where initially all population is on a non-flat band and then the system experiences population transfer to other two bands.

\begin{figure} 
\centering 
\includegraphics[width=\linewidth]{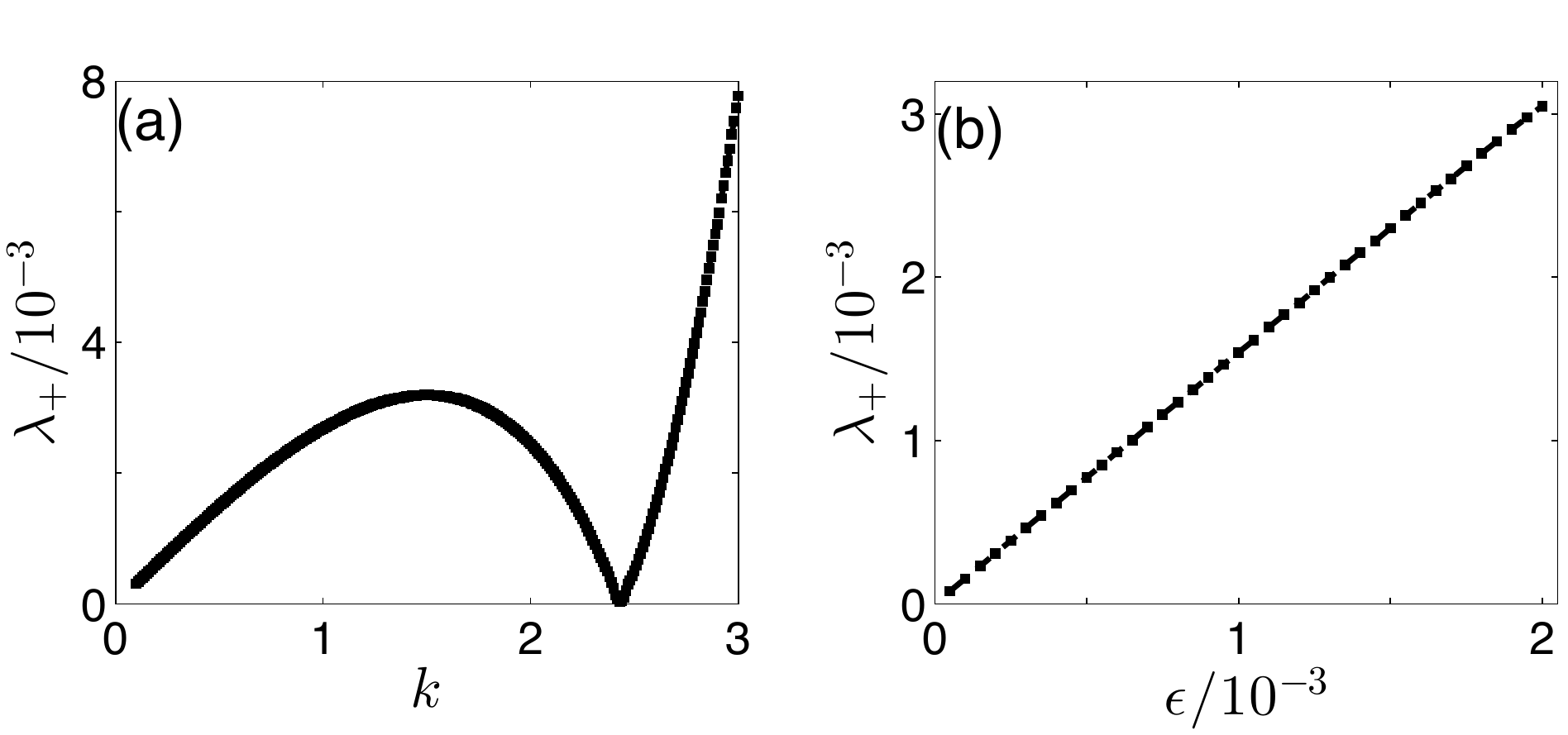}
\caption{Exponential rate of EQS obtained by direct exponential fitting of the time dependence of the kinetic energy (i.e., $\sim e^{2\lambda_+}$) over a proper time scale, with $\tau=2\pi/3+\epsilon$. In panel (a), $\epsilon=2\pi/3003$,  $\lambda_+$ is shown for a varying $k$. In panel (b),  $k=1.5$,  $\lambda_+$ is shown for a varying $\epsilon$. }
\label{fig:rDKRM_Lypunov}
\end{figure}

We have also numerically examined how the exponential rate of EQS depends on the system parameters $k$ and $\epsilon$ for a flat-band initial state, with the results shown in Fig.~\ref{fig:rDKRM_Lypunov}.  In particular, we first fit the EQS behavior shown in Fig.~\ref{fig:rDKRM_evo_1_3_k_2}(a) by $e^{2\lambda_+}$ over a proper time scale, and then plot the obtained $\lambda_+$ vs $k$ [Fig.~\ref{fig:rDKRM_Lypunov}(a)] or vs $\epsilon$ [Fig.~\ref{fig:rDKRM_Lypunov}(b)].  It is seen that the exponential rate is a linear function of $\epsilon$, but displays a highly nontrivial dependence on $k$.  As such,   the exponential rate of EQS depends on both parameters $k$ and $\epsilon$, rather than depending on their product only.  This marks a clear difference from the analysis in Ref.~\cite{GongJB2011} for $\tau=2\pi+\epsilon$.

Parallel numerical studies are also carried out for other cases.  For example, we have considered cases with much smaller values of $k$.  Due to the above-mentioned power-law scaling of the band width ($\sim k^{q+2}$) for $\udkrc$ with both $p$ and $q$ being odd, all the Floquet bands are effectively flat. In these situations (including the high-resonance case studied in Ref.~\cite{GongJB2011}), we find that upon introducing a detuning of $\epsilon$, {\it i.e.} $\tau=2\pi p/q+\epsilon$, EQS occurs and the result is insensitive to the initial state preparation. This further strengthens the view that the flat-band initial states are important to understand EQS. To double-check this, we have also considered cases in which either $p$ or $q$ is an even integer. In these cases there are no flat bands for  $\udkrc$ and indeed we do not find EQS either. All these numerical results suggest an important connection between EQS and the existence of a flat (or effectively flat) band in $\udkrc$ with both $p$ and $q$ being odd.

\subsection{Theoretical analysis}
\subsubsection{Single-Band Approximation.}
To see how a detuning from exact resonance ($\tau=2\pi p/q+\epsilon$) induces nontrivial dynamical evolution,  we first rewrite the ORDKR propagator in the following form:
\begin{equation}
\udkrb = \hat{R}_{\epsilon}^{\dagger} {\hat W}_{\tau} {\hat R}_{\epsilon} \;,
\label{dec1}
\end{equation}
where
\begin{equation}
\label{dec2}
\hat R_{\epsilon} = \exp(-i\epsilon{\hat l}^2/2) \;,
\end{equation}
\begin{equation}
{\hat W}_{\tau} = \udkrc{\hat U}_{\epsilon} \;,
\end{equation}
and so
\begin{equation}
{\hat U}_{\epsilon} = \exp[ik\cos(\hat\theta)]{\hat R}_{\epsilon}\exp[-ik\cos(\hat\theta)]{\hat R}^{\dagger}_{\epsilon} \;.
\label{dec3}
\end{equation}
Operator $\hat W_{\tau}$ is thus a product of the exactly resonant  propagator times a detuning-induced factor. This factor  $\hat U_{\epsilon}$ has two effects: it breaks the momentum-space translational invariance possessed by $\udkrc$, which means that quasi-position is no longer a conserved quantity; moreover, it causes population transfer between different Floquet bands of $\udkrc$.

Our analysis is based on two main ingredients. First,  we will implement an approximation of the Born-Oppenheimer type, motivated by our numerical results discussed above, that the population transfer can be insignificant over a considerable time scale for sufficiently small detuning. We shall therefore project the above operators  onto the band subspaces, thus neglecting inter-band interactions \cite{Guarneri2009}. Next, to study  the single-band dynamics thus obtained at small detuning $\epsilon$  we will implement a pseudo-classical approximation, resting on the observation that in the dynamical equations $\epsilon$ plays the formal role of a Planck constant. The pseudo-classical approximation \cite{classical} is  based on the limit $\epsilon\to 0$, $\epsilon l\to I$, and $k\epsilon\to\tilde k$; where $I$ is a pseudoclassical momentum, that is conjugate to position $\theta$, when $\epsilon$ is given the role of a Planck constant.

\begin{figure} 
\centering 
\includegraphics[width=\linewidth]{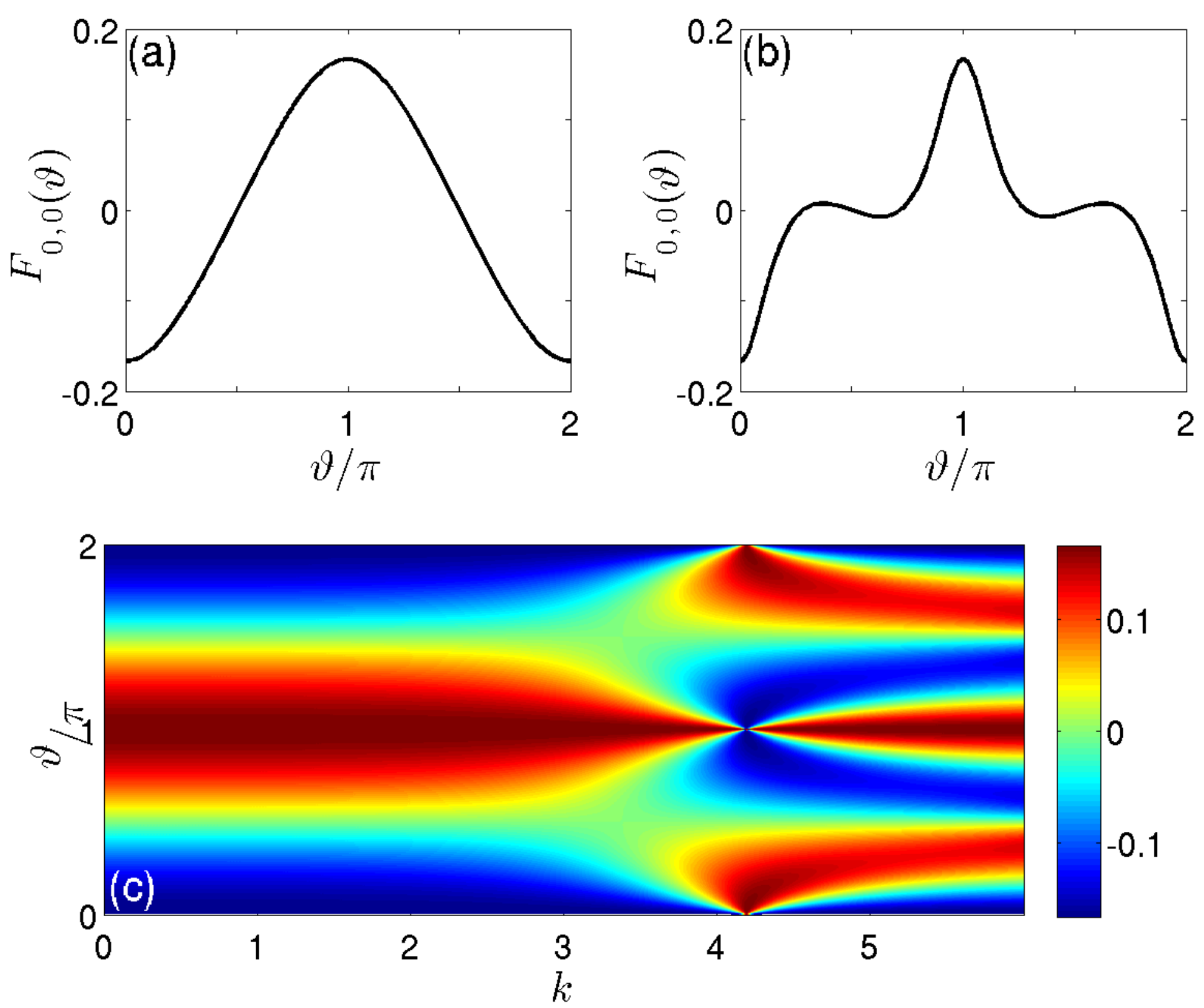}
\caption{ (color online) Effective potential for flat Floquet band in spinor representation, with $\tau=2\pi/3$. In panel (a), $k=2.0$, and in panel (b),  $k=3.5$. Panel (c) is the top view of the effective potential with respect to $\vartheta$ and $k$. The color bar indicates the value of $F_{0,0}(\vartheta)$. }
\label{fig:effective_band_potential}
\end{figure} 

To project onto band subspaces we have first to write operators in Eq.~(\ref{dec2}) and (\ref{dec3}) in the band representation.  Matrix elements of operator $\hat R_{\epsilon}$ in Eq.~(\ref{dec2}) are computed using Eq.~(\ref{varthetanu}). The task is simplified by the observation  that in the pseudoclassical approximation $\epsilon\to 0$, to be implemented later, terms like $\epsilon\tilde l$ and $\epsilon\tilde l^2$ will be negligible, because $\tilde l$ is bounded by $q$. Dismissing such terms, we find :
\begin{equation}
\langle\vartheta,\nu|\hat R_{\epsilon}|\vartheta',\nu'\rangle=\delta_{\nu,\nu'}\langle\vartheta,\nu|\exp(-i\frac{q^2\hat L^2}{2\epsilon})|\vartheta',\nu\rangle \;,
\label{rmatel}
\end{equation}
where $\hat L\equiv -i\epsilon\frac{d}{d\vartheta}$ is the operator that is pseudoclassically conjugate to quasi-position. Thus, in the pseudoclassical limit, operator $\hat R_{\epsilon}$ does not cause interband transitions. The calculation of matrix elements of the kick operator is slightly more complicated. For an arbitrary function $G(\hat\theta)$ of the position operator, Eq.(\ref{varthetanu}) yields:
\begin{equation}
\begin{split}
&\langle\vartheta,\nu|G(\hat\theta)|\vartheta',\nu'\rangle \\
=& \int_0^{2\pi}d\theta G(\theta)\langle\vartheta,\nu|\theta\rangle\langle\theta|\vartheta',\nu'\rangle \\
=& \delta(\vartheta-\vartheta')\sum\limits_{r=0}^{q-1}V^*_{\nu,r}(\vartheta)V_{\nu',r}(\vartheta)G(\vartheta/q+2\pi r/q) \;,
\end{split}
\label{kmatel1}
\end{equation}
where:
\begin{equation}
V_{\nu,r}(\vartheta) = \frac{1}{q}\sum\limits_{\tilde l=0}^{q-1}u_{\tilde l}^{(\nu)}(\vartheta) \exp\left[i\tilde l(\vartheta/q+2\pi r/q)\right] \;.
\end{equation}
With $G(\theta)=\cos(\theta)$, this yields
\begin{equation}
\langle\vartheta,\nu|\cos(\hat\theta)|\vartheta',\nu'\rangle = \delta(\vartheta-\vartheta')F_{\nu,\nu'}(\vartheta) \;,
\end{equation}
where:
\begin{equation}
F_{\nu,\nu'}(\vartheta) = \frac{1}{2q}\sum\limits_{\tilde l=0}^{q-1}\left[u_{\tilde l}^{(\nu)}(\vartheta)u_{\tilde l+1}^{(\nu')*}+u_{\tilde l}^{(\nu')}(\vartheta)u_{\tilde l-1}^{(\nu')*}\right] \;.
\label{fnunu}
\end{equation}
where $\tilde l\pm 1$ is understood mod$(q)$. This shows that the kick operators $\exp[\pm ik\cos(\hat\theta)]$ do produce interband transitions, even in the pseudoclassical limit. In order to obtain a projected single-band dynamics  without breaking unitarity, we simply ignore  all interband matrix elements of $\cos(\hat\theta)$. For the $\nu$-th band this amounts to replacing operator $\hat U_{\epsilon}$ by the Born-Oppenheimer-like operator:
\begin{equation}
\begin{split}
\hat U^{(\text{\tiny{BO}})}_{\nu,\epsilon} &=  \exp\left(\tfrac{i}{\epsilon}\tilde kF_{\nu,\nu}(\vartheta)\right) \exp\left(-\frac{i}{\epsilon}q^2\hat L^2\right) \\
&\times \exp\left(-\frac{i}{\epsilon}\tilde kF_{\nu,\nu}(\vartheta)\right) \exp\left(\tfrac{i}{\epsilon}q^2\hat L^2\right) \;.
\end{split}
\label{bop}
\end{equation}
It should be noted that the ``band potential" $F_{\nu,\nu}(\vartheta)$ is gauge-invariant, {\it i.e.} it does not depend on the choice of the arbitrary phase factor of the band eigenvectors. For higher ($q>1$) resonances it has to be numerically computed. In the case $p=1$, $q=3$, the effective potentials for all bands can be found analytically. We plot in Fig.~\ref{fig:effective_band_potential} the effective potential for flat band. The flat-band potential $F_{0,0}$ is found to be indistinguishable from $-(2q)^{-1}\cos(\vartheta)$ when $k<<1$.

\subsubsection{Pseudoclassical approximation.}

\begin{figure} 
\centering 
\includegraphics[width=\linewidth]{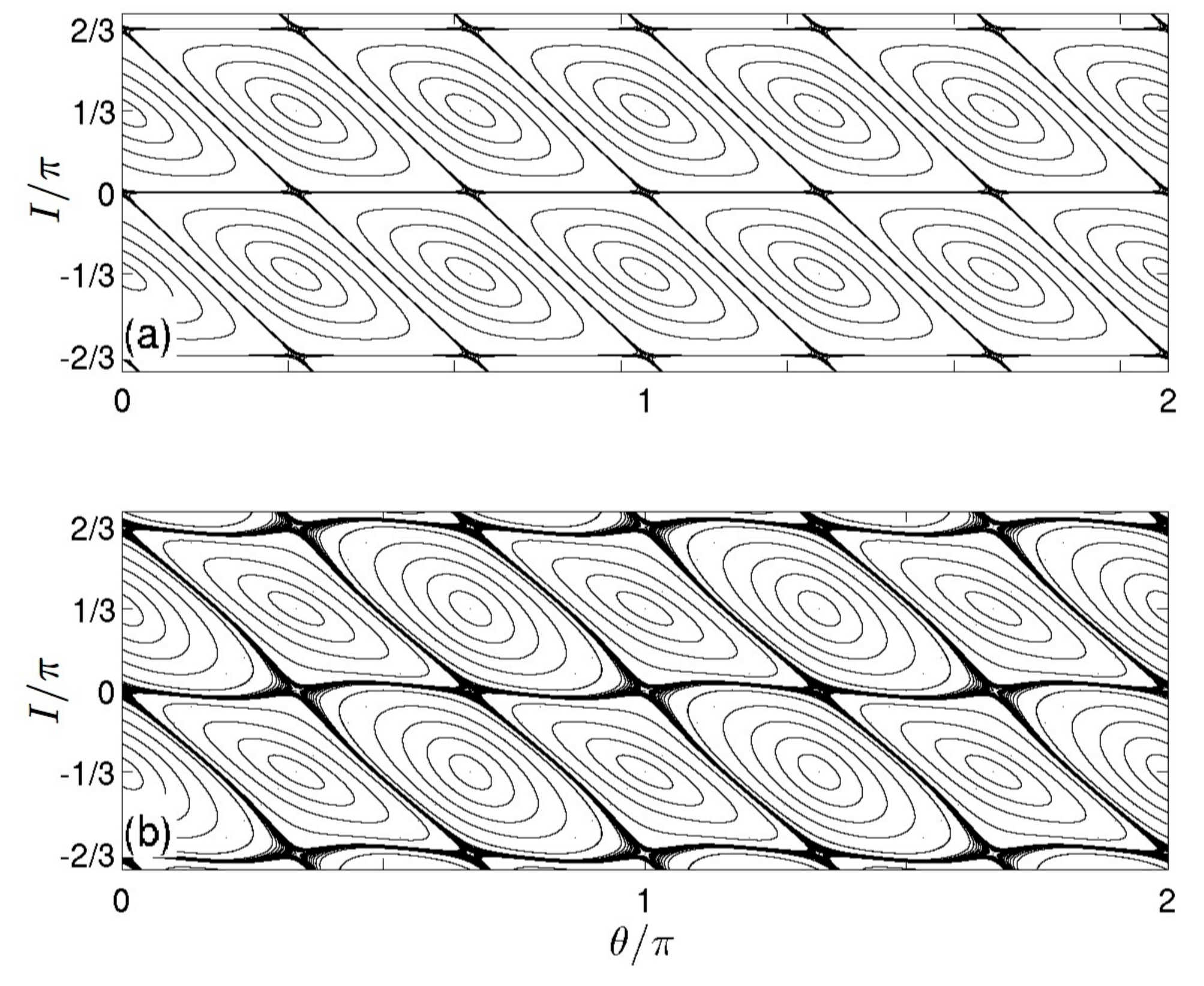}
\caption{(a) Phase-space portrait of the pseudoclassical map of ORDKR described by (\ref{twomaps}), for flat band with $k=2.0$, $\epsilon=10^{-3}$, and $\tau=2\pi/3 +\epsilon$.  (b) Phase-space portrait of the pseudoclassical map described in (\ref{fourmaps2}) for nonflat band with $k=2.0$, $\epsilon=10^{-3}$, and $\tau=2\pi/3 +\epsilon$. }
\label{fig:phasespace}
\end{figure}

\begin{figure} 
\centering 
\includegraphics[width=\linewidth]{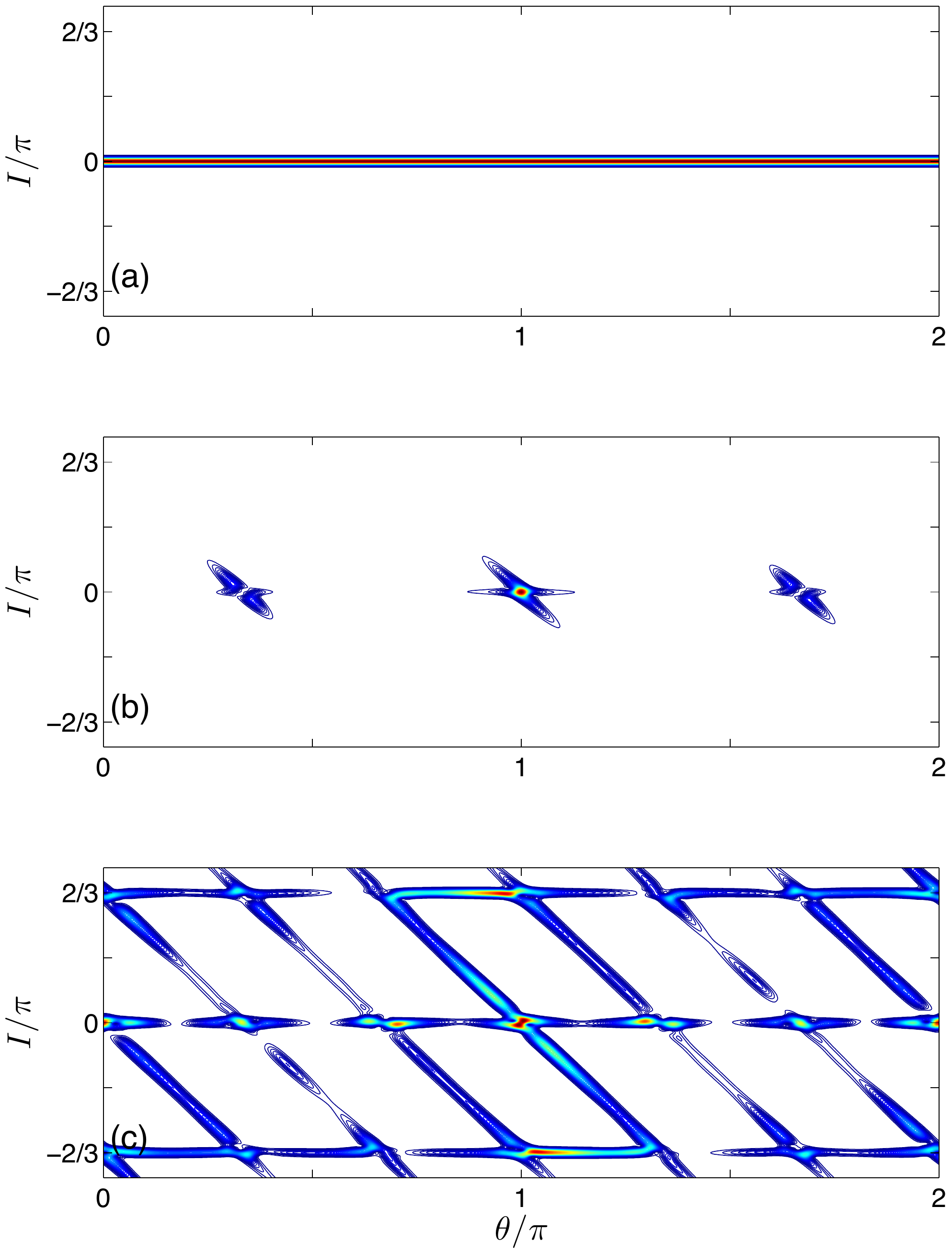}
\caption{ (color online) Husimi distribution for $\udkrc$ ($p=1, q=3$), with initial state $I=0$ under condition $k=2.0$ and $\tau=2\pi/3+2\pi/3003$. (a) Husimi distribution of the initial state.  (b) Husimi distribution of the state after 3000 iterations under the map $\udkrb$.  (c) Husimi distribution of the state after 60000 iterations under the map $\udkrb$. In plotting the Husimi distribution
the dimensionless effective Planck constant used in the coherent states is taken to be
the same as the detuning $2\pi/3003$. }
\label{fig:husimi_a}
\end{figure}

If $\epsilon$ is regarded as a (pseudo)-Planck constant, then Eq.~(\ref{bop}) is manifestly the formal quantization of the  pseudoclassical map $(\vartheta_0,L_0)\mapsto(\vartheta_4,L_4)$, that is obtained by composing the following four maps:
\begin{equation}
\begin{split}
(\vartheta_1,L_1) &= (\vartheta_0-q^2L_0,L_0) \\
(\vartheta_2,L_2) &= (\vartheta_1,L_1-\tilde kF'_{\nu,\nu}(\vartheta_1) \\
(\vartheta_3,L_3) &= (\vartheta_2+q^2L_2,L_2) \\
(\vartheta_4,L_4) &= (\vartheta_3,L_3+\tilde kF'_{\nu,\nu}(\vartheta_3)) \;.
\end{split}
\label{fourmaps1}
\end{equation}
These maps respectively correspond to the four unitary operators of which Eq.~(\ref{bop}) is composed. Using this, and Eq.~(\ref{dec3}), we can then  derive a pseudoclassical in-band approximation for the operator $\hat W_{\tau}$.
In the $\nu$-th band, $\udkrc$ is just multiplication by $\exp(-i\Omega(\vartheta))$, which  amounts to an additional kick, implies a correction by:
\begin{equation}
L_4 \rightarrow L_4+\epsilon\Omega'(\vartheta) \;.
\label{correc}
\end{equation}
The additional operators $\hat R_{\epsilon}$,$\hat R^{\dagger}_{\epsilon}$ which appear in Eq.~(\ref{dec1}) simply enforce a cyclic rearrangement of the maps in Eq.~(\ref{fourmaps1}), such that the 2nd map in Eq.~(\ref{fourmaps1}) becomes the 1st, while  the 1st becomes the 4th. This composite map is however written in the ``band" variables $\vartheta,L$, and to make  comparisons to the exact dynamics it is necessary to restore the ``physical" variables $\theta$ and $I=\epsilon l$ . To this end  we use that $\vartheta=q\theta$, and so $I=qL$. This yields:
\begin{equation}
\begin{split}
(\theta_1,I_1) &= (\theta_0,I_0-q\tilde kF'_{\nu,\nu}(q\theta_0)) \\
(\theta_2,I_2) &= (\theta_1+I_1,I_1) \\
(\theta_3,I_3) &= (\theta_2,I_2+q\tilde kF'_{\nu,\nu}(q\theta_2) \\
I_3 &= I_3+\epsilon q\Omega'(q\theta_3) \\
(\theta_4,I_4) &= (\theta_3-I_3,I_3) \\
\end{split}
\label{fourmaps2}
\end{equation}
In the case of the flat band ($\nu=0$), the correction in Eq.~(\ref{correc}) is absent; hence the pseudoclassical analog of the complete operator $\udkrb$ is found, once more as the product of four maps.
The fourfold product of maps in the variables $\theta,I$ which is obtained in this way is our final pseudoclassical approximation for the flat-band dynamics. Its form is greatly simplified by one more  canonical change of variables, from $\theta,I$ to $\theta,J\equiv\theta+I$. In the new variables, the map can be written in an alternate form as iteration map:
\begin{equation}
\begin{split}
J_{n+1} &=J_n-B(\theta_n)\\
\theta_{n+1} &=\theta_n-B(J_{n+1})\\
\end{split}
\label{twomaps}
\end{equation}
where $B(\theta)=q\tilde kF'_{0,0}(q\theta)$ and $B(J)=q\tilde kF'_{0,0}(qJ)$.
 This  map is somewhat resemblant of  the Kicked-Harper (KH) map, from which it differs because in that case $B(\theta)\propto\sin(\theta)$. It is important to remark that this map is not precisely pseudoclassical, because along with the pseudoclassical parameter $\tilde k$ it still explicitly retains $\epsilon$, the pseudo-Planck constant. Note also that the band potential is fully  determined by the spectral structure at exact resonance, so it is only dependent  on $k={\tilde k}/\epsilon$. At small $\epsilon\neq 0$ the phase space structure of the map, illustrated in Fig.~\ref{fig:phasespace}(a) for the case $q=3$,  reveals the origin of the EQS, as we shall explain in detail. For not too large $k$, the phase portrait exhibits $2q$ unstable fixed points in each line $I=2n\pi/q$ ($n$ an arbitrary integer). The stable and unstable manifolds of such points support thin stochastic layers, that interconnect so as to form a regular network, whereby phase space is partitioned in parallelograms. Each parallelogram is centered at a stable fixed point and has unstable fixed points at its vertices; two of its sides  are parallel to the $\theta$-axis, and the other two are at an angle of $\arctan(-2)$ with it.  For half of the unstable points which lie on the zero-momentum axis, the unstable manifolds reach out in momentum space towards the unstable points located at momentum $\pm 2\pi/q$. Hence, a pseudoclassical ensemble of points initially prepared near momentum zero will be attracted exponentially fast along the zero momentum axis by such fixed points, to be then  exponentially fast driven away from the axis, along their  unstable manifolds. The same qualitative behavior is observed  in the dynamics of quantum states, that are initially prepared in the flat band subspace of $\udkrc$, near zero momentum. For such states, the quantum evolution follows for a long time the flat-band dynamics as we have seen, and the small-$\epsilon$, quantum flat-band dynamics in turn mirrors the pseudoclassical dynamics. This is confirmed in Fig.~\ref{fig:husimi_a}, where  Husimi phase-space distributions are shown for a quantum state evolved from an initial state that prepared on the flat band of $\udkrc$.
The pseudoclassical separatrix structure is faithfully reflected in the quantum dynamics.
 Because the actual physical momentum is given by $\tau I/\epsilon$,  the EQS in the expectation value of $I^2$ results in a large-scale EQS in the actual momentum space.  This two-fold EQS mechanism explains why the system should be slightly detuned from $\tau=2\pi p/q$ with $p$ and $q$ both being odd. If either $p$ or $q$ is an even number, then $\udkrc$ does not have a flat band (or an effectively flat band state) and our analysis does not apply.

To further check the above picture, we consider an initial state prepared on a non-flat band of $\udkrc$ with $p=1$, $q=3$ and $k=2$ (we do not consider a smaller $k$ because, as mentioned earlier, all bands will be effectively flat if $k$ is very small).  In this case, $\Omega'(\vartheta)$ is not negligible, and hence an appropriate classical dynamics description should be based on the map in Eq.~(\ref{fourmaps2}). More importantly, which is perhaps not obvious from the phase space plot in Fig.~\ref{fig:phasespace}(b), along one individual trajectory, the value of $L_\epsilon$ can now jump by $\epsilon q\Omega'(\vartheta)$, which yields a linear increase in the momentum scale (on average) and has nothing to do with an exponential repulsion mechanism we identified earlier.
Indeed,  the computational results shown in Fig.~\ref{fig:rDKRM_evo_1_3_k_2}(b) do not suggest any EQS behavior. Rather, examining the results using a log-log plot shows that the spreading is ballistic, which is consistent with the fact that the states are prepared on a continuous non-flat band.

\subsubsection{Quantitative Investigation of EQS Rates}

\begin{figure} 
\centering 
\includegraphics[width=\linewidth]{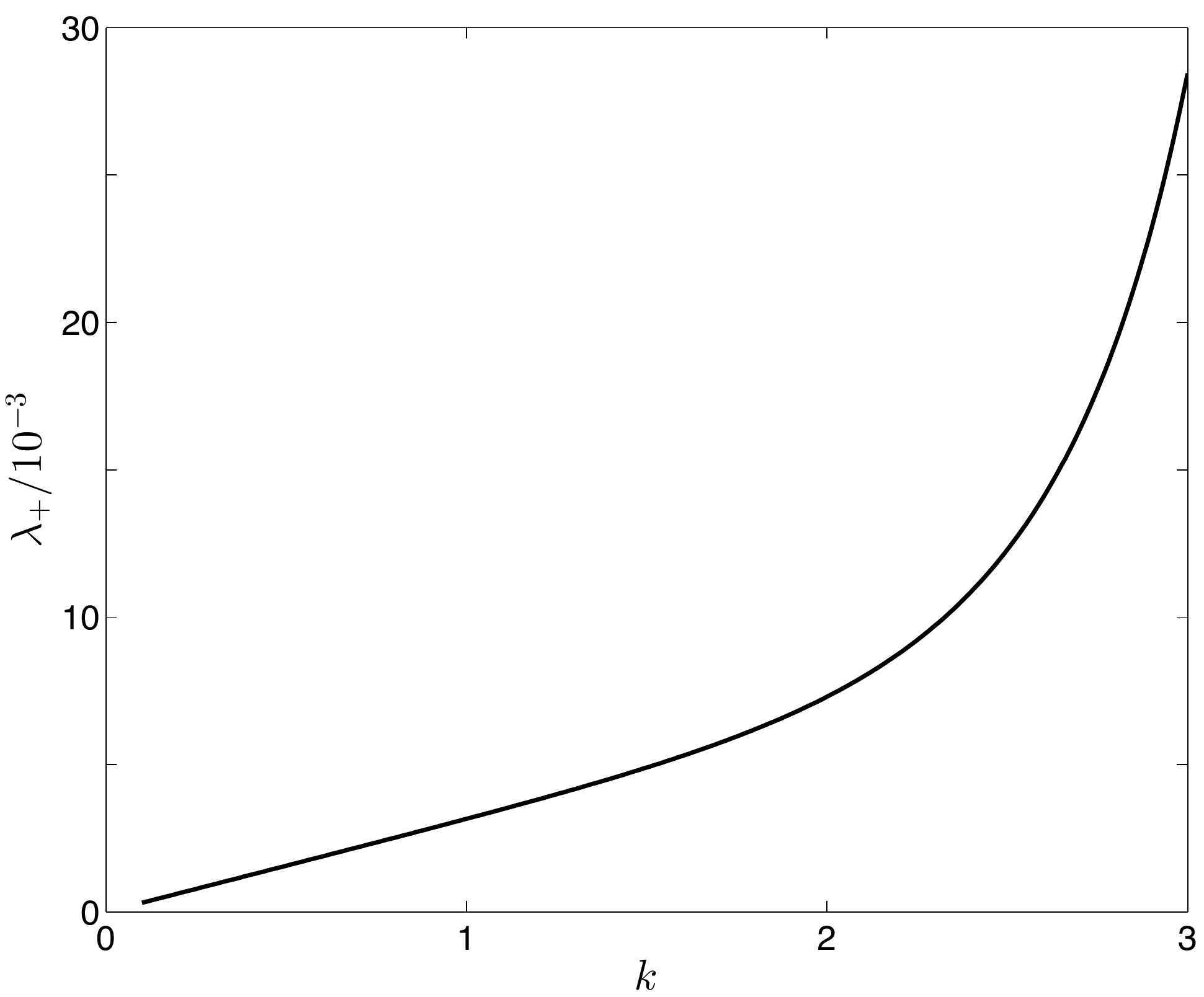}
\caption{Pseudoclassocal prediction of the EQS rate at the unstable fixed point for our pseudo-classical map, with $\tau=2\pi/3+ \epsilon$, $\epsilon=2\pi/3003$.}
\label{fig:rDKRM_Lypunov-cl}
\end{figure} 

To perform linear stability analysis of the pseudoclassical map described in Eq.~(\ref{twomaps}), we wrote the Jacobian matrix (as the stability matrix) at the fixed points
\begin{equation}
\JM(\theta,J)=
\begin{pmatrix}
1	&-B'(\theta)\\
-B'(J)&1+B'(\theta)B'(J)\\
\end{pmatrix}
\end{equation}
where $J\equiv \theta+I$, $B(\theta)=q\tilde kF'_{0,0}(q\theta)$, and $B(J)=q\tilde kF'_{0,0}(qJ)$.
 The unstable points can be obtained from the band potential derived above.
It can be easily shown that the pseudoclassical map is an area-preserving map, as $\det{(\JM)}\equiv 1$. Based on a linear stability analysis at the unstable fixed points of the pseudo-classical map, we immediately obtain a pseudo-classical prediction of the EQS rate, i.e., $\lambda_{\pm}^{cl} = \ln[\frac{2+K\pm\sqrt{K^2+4K}}{2}]$ with $K=B'(\theta)B'(J)$ evaluated at an unstable fixed points.  For small values of kicking strength $k$,
$K$ is proportional to $k^2$ and $\lambda_{+}^{cl}\sim k$.
In Fig.~\ref{fig:rDKRM_Lypunov-cl}, we show $\lambda_{+}^{cl}$ vs $k$. Note that if we use an actual ensemble of trajectories close to zero momentum to simulate the pseudo-classical dynamics, the obtained pseudo-classical EQS rate still agrees quite well with the simple linear stability analysis here.


Interestingly, our numerical results shown in Fig.\ref{fig:rDKRM_Lypunov} are richer than the prediction of $\lambda_{+}^{cl}\sim k$:
 the actual EQS rate $\lambda_{+}$ is found to be proportional to $\epsilon$ but is not a monotonous function of $k$.  This disagreement on the quantitative level suggests that our above treatments have introduced some errors.  Additional numerical checks show that the errors are not directly due to the pseudo-classical treatment itself.

\begin{figure} 
\centering 
\includegraphics[width=\linewidth]{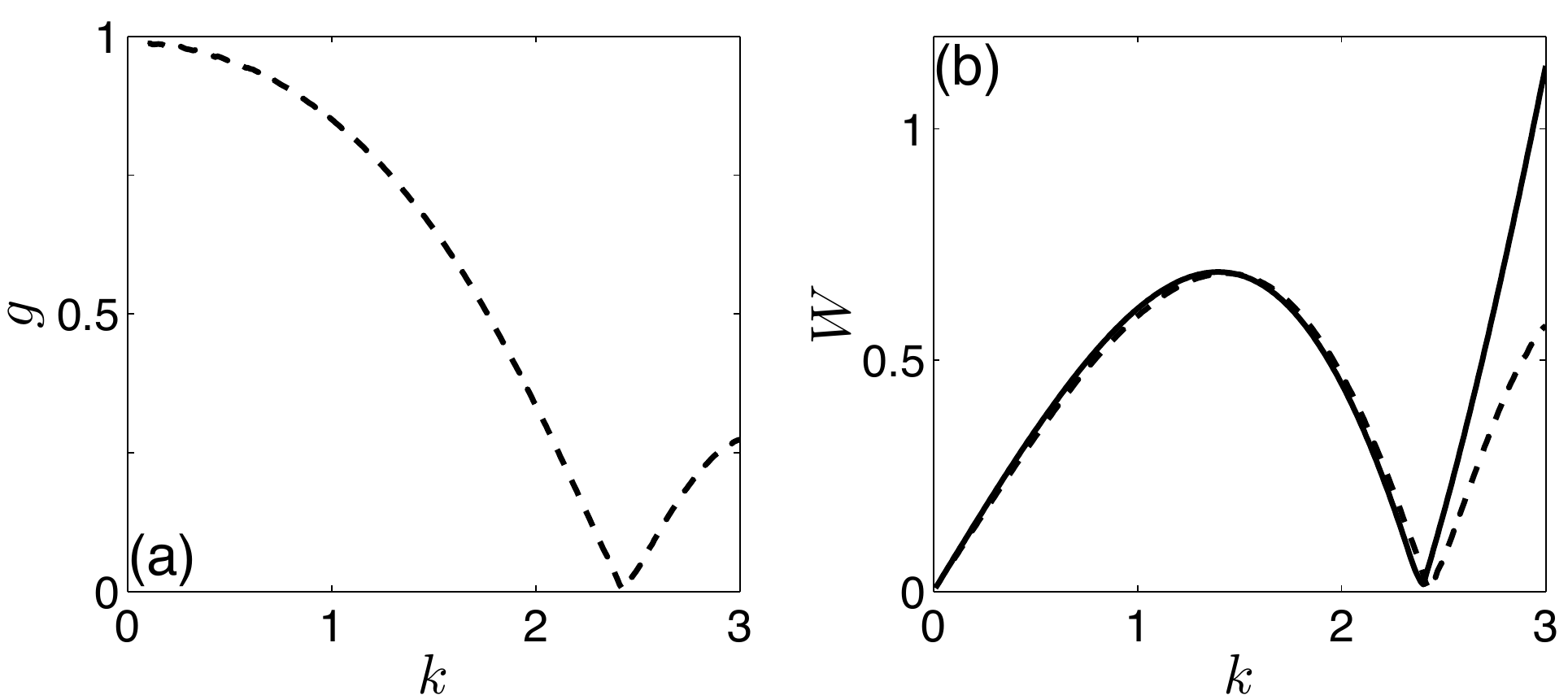}
\caption{(a) Dashed line represents numerical results of $g$, where $g$ is defined in the text as the ratio of the actual exponential rate and the rate obtained from our pseudo-classical map, with $\tau=2\pi/3+ \epsilon$, $\epsilon=2\pi/3003$.  In panel(b) the dashed line basically present the same result but plotting $0.7\times g\times k$ for a varying $k$. Solid line represents $W(k)$, which is the spectral range of the middle subband cluster of $\udkrb$ (see Fig. 2).}
\label{fig:rDKRM_Lypunov-g_factor}
\end{figure} 

\begin{figure} 
\centering 
\includegraphics[width=\linewidth]{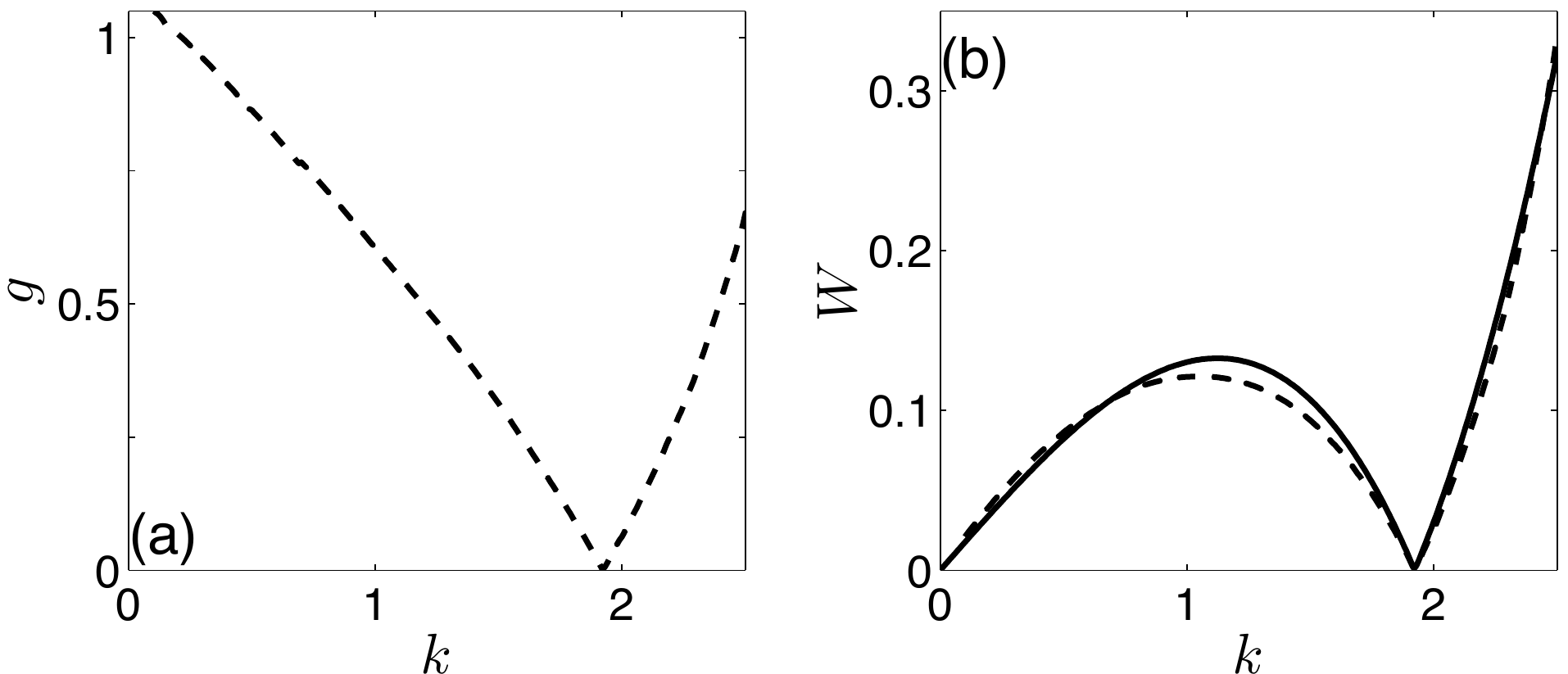}
\caption{(a) Dashed line represents numerical results of $g$, where $g$ is defined in the text as the ratio of the actual exponential rate and the rate obtained from our pseudo-classical map, with $\tau=6\pi/5+ \epsilon$, $\epsilon=2\pi/4990$.  In panel(b) dashed line basically presents the same result but plotting $0.2\times g\times k$ for a varying $k$. Solid line represents $W(k)$, which is the spectral range of the middle subband cluster of $\udkrb$.}
\label{fig:rDKRM_Lypunov-g_factor_2}
\end{figure} 

To better characterize and understand the quantitative disagreement we introduce a factor $g\equiv \frac{\lambda_{+}}{\lambda_{+}^{cl}}$, namely, the ratio of the numerical EQS rate and that predicted by our pseudo-classical map.  $g$ thus defined should be a function of $\epsilon$ and $k$. The dependence of $g$ on $\epsilon$ is found to be very weak so we focus on the $k$-dependence. In Fig.~\ref{fig:rDKRM_Lypunov-g_factor}(a), we show $g$ vs $k$. For the same detuning, Fig.~\ref{fig:rDKRM_Lypunov-g_factor}(b) shows $0.7\times g\times k$ vs $k$ (dash line) as compared with $W(k)$, where $W(k)$ is the spectral range of the middle subband cluster of $\udkrb$, with $\tau=2\pi/3+ 2\pi/3003$. Remarkably, the $g$ factor is seen to be strongly correlated with the actual spectral range $W(k)$ of the middle subband cluster [for a computational example of the subband clusters, see Fig.~\ref{fig:rDKRM_Flo_spectrum}(b)]. This correlation between $g$ and $W(k)$ is somewhat expected: the pseudoclassical Hamiltonian suggests that the energy scale is proportional to $k$ but the actual spectrum can be a highly nonlinear function of $k$. In particular, it is seen that for $k\approx 2.4$, $g$ as well as the spectral width $W(k)$ are seen to be almost zero in Fig.~\ref{fig:rDKRM_Lypunov-g_factor}(a). This collapse of the Floquet subbands leads to a freezing of the quantum dynamics and the difference between the the actual dynamics and our pseudoclassical prediction becomes most pronounced. Such type of information about $\udkrb$ can be sensitive to $\tau$ and $k$, and is naturally not considered in our above theoretical analysis based on the adiabatic approximation and the pseudo-classical treatment. Roughly speaking, detailed aspects of the dynamical evolution are determined by the actual spectrum of $\udkrb$, not by the on-resonance propagator $\udkrc$.  So by analyzing the actual dynamics using one band subspace of $\udkrc$ only, certain subtle quantum effects connected with the many subbands are necessarily lost. Indeed, in our theoretical analysis, we only require the population to stay on the flat band of $\udkrc$ and have neglected all possible fine structure in the actual Floquet spectrum of $\udkrb$.

Some additional remarks are in order.  First, we have also carried out the same analysis for $\tau=2\pi+\epsilon$, i.e., the anti-resonance case studied in Ref.~\cite{GongJB2011}.  In this case we always find $g=1$.   That is, for ORDKR slightly detuned from an anti-resonance condition, the pseudo-classical prediction is found to match the EQS rate quantitatively.  We believe that this is closely connected with the fact that $\udkrc$ ($p=1$, $q=1$) has only one band (which is flat) and hence there is no need to apply the above-mentioned adiabatic approximation. We have also studied other cases, for example, with $\tau=6\pi/5+\epsilon$ and $\epsilon=2\pi/4995$, see Fig.~\ref{fig:rDKRM_Lypunov-g_factor_2}, again showing correlations with the actual spectral range of the subband clusters of $\udkrb$.  As the focus of this work is on a physical explanation of EQS in ORDKR detuned slightly from a general resonance condition, which has been achieved, we leave a more detailed study of $g$ for possible future work.


\section{Concluding remarks}

The main contribution of this work is to extend the analysis in Ref.~\cite{GongJB2011} from ORDKR near an antiresonance case to general cases near high-order resonances. It is explained here why EQS can occur in ORDKR if its effective Planck constant $\tau$ is slightly detuned from $2\pi p/q$ with both $p$ and $q$ being odd. Our theoretical analysis shows that EQS is closely related to two pieces of physics: (i) the existence of flat bands or effectively flat bands of ORDKR under high-order resonances and (ii) the emergence of an integrable pseudo-classical limit whose dynamics can induce rather uniform exponential spreading in the momentum space.  We also point out that the pseudo-classical picture also makes it straightforward to understand the time scale of EQS \cite{GongJB2011}. On the quantitative level,  we find that there is some difference in EQS rates between our simple theoretical analysis and the actual quantum dynamics. This indicates that the dynamics of ORDKR near high-order resonances can be richer if details become important.

In this paper we do not discuss any experimental issues.  For example, in atom-optics realizations of the kicked rotor dynamics \cite{recentKR}, the nonzero width in the quasi-momentum should introduce more complications, but in our constructed ORDKR the width is taken as absolutely zero. However, for near anti-resonance cases  this issue was already carefully addressed by us in Ref.~\cite{GongJB2011}.  Therefore we do not repeat similar analysis here.

Long-lasting EQS is an intriguing dynamical phenomenon.  Our route towards the understanding of EQS also indicates that a pseudo-classical approach of kicked systems near quantum resonance conditions constitutes a powerful tool in digesting quantum dynamics that is nevertheless in the deep quantum regime. Further studies on the quantitative difference between our pseudoclassical predictions and the actual quantum results might bring deeper understandings of this approach as well as the adiabatic approximation we made. The very existence of flat Floquet bands of ORDKR and its role in generating EQS also hint that there should be more interesting physics to be discovered in kicked-rotor systems.

\vspace{0.5cm}
{\bf Acknowledgments:} J.G. is supported by grant ARF Tier I, MOE of Singapore (Grant No. R-144-000-276-112). J.W. received support from NNSF (Grant No.11275159) and SRFDP (Grant No. 20100121110021) of China.


\appendix
\section{Resonant Floquet operators in the spinor representation.}
All Floquet operators considered in this paper may be expressed as products of  unitary  $\hat M$ and $\hat V$ operators. Consequently, their matrix elements in the spinor $(\vartheta,\tilde l)$-representation can be found from matrix elements of the latter operators. These are explicitly calculated in this Appendix.
\label{matelapp}
\begin{equation}
\langle\vartheta,\tilde l|\hat M_{\tau}|\vartheta', \tilde l'\rangle = \sum_{l=-\infty}^{+\infty} \langle\vartheta,\tilde l|l\rangle\langle l|\vartheta',\tilde l'\rangle e^{-i\tau l^2/2}.
\end{equation}
Using Eq.~(\ref{eigfs}), replacing $\tau=2\pi p/q$, $l=\overline lq+\tilde l''$, and summing over $\overline l,\tilde l''$, one finds:
\begin{equation}
\langle\vartheta,\tilde l|\hat M_{\tau}|\vartheta', \tilde l'\rangle = \delta(\vartheta-\vartheta'-\pi p q){\cal M}_{\tilde l\tilde l'}
\end{equation}
where the matrix ${\cal M}_{\tilde l\tilde l'}=\delta_{\tilde l,\tilde l'}\exp(-i\pi p\tilde l^2/2)$.

In a completely similar manner one finds:
\begin{equation}
 \langle\vartheta,\tilde l|\hat V|\vartheta', \tilde l'\rangle = \delta(\vartheta-\vartheta'){\cal V}_{\tilde l\tilde l'}(\vartheta)
\end{equation}
where:
\begin{equation}
{\cal V}_{\tilde l\tilde l'}(\vartheta) = \frac{1}{q}\sum_{j=0}^{q-1} e^{-ik\cos(2\pi j/q+\vartheta/q)} e^{-i(\tilde l-\tilde l')(2\pi j/q+\vartheta/q)} \;.
\end{equation}

For $\tau=2\pi p/q$ with both $p$ and $q$ being odd integer, we have
\begin{equation}
\begin{split}
&\caludkrb=e^{i\tau\tilde l^2 /2}\sum_{\tilde l''=0}^{q-1} e^{-i\tau\tilde l''^2 /2} e^{i\pi (\tilde l-\tilde l'')}\\
&\frac{1}{q}\sum_{j_1=0}^{q-1}e^{ i k_1 \cos(2\pi j_1/q+\vartheta/q)}e^{i (2\pi j_1/q+\vartheta/q)(\tilde l'-\tilde l'')}\\
&\frac{1}{q}\sum_{j_2=0}^{N-1}e^{ -ik_2 \cos(2\pi j_2/q+\vartheta/q)}e^{i (2\pi j_2/q+\vartheta/q)(\tilde l''-\tilde l)}
\end{split}
\end{equation}

\section{A solvable case with $\tau=2\pi/3$}
\label{solvablecase.apx}

For ORDKR  under the resonance condition $\tau=2\pi/3$, there are three Floquet bands and all of them can be analytically solved. The eigenphases are found to be
\begin{equation}
\begin{split}
&\Omega_1=\Omega,\\
&\Omega_0=0,\\
&\Omega_{-1}=-\Omega,
\end{split}
\end{equation}
where
\begin{equation}
\begin{split}
&\Omega=\cos^{-1}(\eta),\\
&\eta=\left (\frac{1}{3}\cos{2\beta}+\frac{2}{3}\cos{\beta}\cos{3\alpha}\right),\\
&\alpha=\frac{k}{2}\cos{\frac{\vartheta}{3}},\\
&\beta=\frac{\sqrt{3}k}{2}\sin{\frac{\vartheta}{3}}.\\
\end{split}
\end{equation}
The eigenphase $\Omega_0$ is independent of the Bloch phase $\vartheta$ and so it gives a perfectly flat band. The corresponding normalized eigenvector is (see sect.\ref{SSP} for notations):
\begin{equation}
\begin{split}
&u_0^{(0)}(\vartheta) =0 \\
u_1^{(0)}(\vartheta) =&\gamma e^{-i\vartheta/3} \left[e^{-3i\alpha}-2\cos(\beta+\pi/3)\right] \\
u_2^{(0)}(\vartheta) =&\gamma e^{-2i\vartheta/3} \left[2\cos(\beta-\pi/3)-e^{-3i\alpha}\right] \\
&\gamma =\left[6(1-\eta)\right]^{-1/2} \;.
\end{split}
\end{equation}

\section{Proof of the Existence of Flat band}
\label{flatb.apx}
Here we give a simple proof that whenever $\tau=2\pi p/q$,  with $p$ and $q$ odd and coprime integers,  $1$ is an infinitely degenerate proper eigenvalue of $\udkrc$, that generates  a flat band in the  band spectrum of $\udkrc$.
From Eq.~(A4) it is immediate that :
$$
{\cal V}(\vartheta+\pi)\;=\;{\cal R}{\cal V}^{\dagger}(\vartheta){\cal R}\;,\;\;\;{\cal R}_{\tilde j\tilde l}:=
\delta_{\tilde j\tilde\l}\;(-1)^{\tilde l}\;,
$$
Hence, defining a matrix ${\cal M}_1:= {\cal R}{\cal M}$, and using that $\cal M$ is diagonal, whenever $pq$ is an odd integer, the factorization, see Eq.~(\ref{dkprod}), may be rewritten as follows:
\begin{equation}
\label{newdkprod}
{\cal U}^{({\text{\tiny
 {ordkr}})}}(\vartheta)\;=\;{\cal M}_1^{\dagger}\;{\cal V}^{\dagger}(\vartheta)\;{\cal M}_1\;{\cal V}(\vartheta)\;.
\end{equation}
Next we introduce the following unitary matrices ${\cal S}(\vartheta)$:
$$
{\cal S}_{\tilde j\tilde l}(\vartheta)\;=\;\;\tfrac 1{\sqrt q}\;e^{i\tilde j\vartheta/q}\;e^{2\pi i p\tilde j\tilde l/q}\;.
$$
For any given $\vartheta$, ${\cal S}(\vartheta)$ defines a rotation of spin axes that carries ${\cal V}(\vartheta)$ to diagonal form. For a generic $q\times q$ matrix-valued function ${\cal A}(\vartheta)$ of quasi-position let  $\check{\cal A}(\vartheta)$ denote the matrix ${\cal S}(\vartheta)^{\dagger}{\cal A}(\vartheta){\cal S}(\vartheta)$. Then (\ref{newdkprod}) yields:
\begin{equation}
\check{\cal U}^{({\text{\tiny
 {ordkr}})}}(\vartheta)\;=\;\check{\cal M}_1^{\dagger}\;\check{\cal V}^{\dagger}(\vartheta)\;\check{\cal M}_1\;\check{\cal V}(\vartheta)\;.
\end{equation}
Straightforward calculation shows that $\check{\cal V}(\vartheta)$ is diagonal, and $\check{\cal M}_1(\vartheta)$ is symmetric. Therefore, the last equation can be rewritten in the form:
\begin{equation}
\label{dkprod3}
\check{\cal U}^{({\text{\tiny
 {ordkr}})}}(\vartheta)\;=\;\check{\cal M}_1^{*}\;\check{\cal V}^{*}(\vartheta)\;\check{\cal M}_1\;\check{\cal V}(\vartheta)\;.
\end{equation}
 where $*$ denotes complex conjugation. It is then immediate that:
 \begin{equation}
 \label{uniteqv}
 (\check{\cal U}^{({\text{\tiny
 {ordkr}})}})^*(\vartheta)\;=\;{\cal W(\vartheta)}\;\check{\cal U}^{({\text{\tiny
 {ordkr}})}}(\vartheta)\;{\cal W^{\dagger}(\vartheta)}\;,
 \end{equation}
 where ${\cal W}(\vartheta)={\cal V}(\vartheta){\cal M}_1$. Eq.~(\ref{uniteqv}) says that the matrix
 $\check{\cal U}^{({\text{\tiny
 {ordkr}})}}(\vartheta)$ is unitarily equivalent to its own complex conjugate, hence its spectrum is invariant under complex conjugation . The same is then true of the spectrum of ${\cal U}^{({\text{\tiny
 {ordkr}})}}(\vartheta)$ , that is indeed unitarily equivalent to $\check{\cal U}^{({\text{\tiny
 {ordkr}})}}(\vartheta)$ by construction. This proves symmetry of the spectrum with respect to the zero eigenphase axis. As the spectrum consists of $q$ points on the unit circle (counting multiplicities), and $q$ is odd, this symmetry entails  that an odd number of eigenvalues must be real, hence equal to $\pm 1$; on the other hand, from (\ref{dkprod3}), det$(\check{\cal U}^{({\text{\tiny
 {ordkr}})}}(\vartheta))=|$det$(\check{\cal M}_1(\vartheta)|^2|$det$(\check{\cal V}(\vartheta))|^2=1$, so at most an even number of eigenvales may be equal to $-1$. So $1$ is always an eigenvalue, independent of $\theta$, and this produces a flat band at zero eigenphase.


\end{document}